%% file: patrizi.tex
\documentclass[copyright,creativecommons]{eptcs}
\providecommand{\event}{}
\providecommand{\volume}{2}
\providecommand{\anno}{2009}
\providecommand{\firstpage}{37}
\usepackage{breakurl}

\usepackage{latexsym}
\usepackage{amssymb}
\usepackage{amsmath}
\usepackage[dvips]{graphicx}
\usepackage{color}
\usepackage{enumerate}
\usepackage{multirow}
\usepackage{psfrag}
\usepackage{pstricks, pst-node, pst-tree}
\usepackage{subfigure}
\usepackage{xspace}

\input{macros}

\let\psgrid\relax

\title{An Introduction to Simulation-based Techniques for Automated Service Composition}
\author{Fabio Patrizi
	\institute{Dipartimento di Informatica e Sistemistica ``A. Ruberti''\\ {\sc Sapienza} Universit\`a di Roma}
\email{patrizi@dis.uniroma1.it}
}
\def\titlerunning{An Introduction to Simulation-based Techniques for Automated Service Composition}
\def\authorrunning{F.~Patrizi}
\begin{document}
\maketitle

\begin{abstract}
This work is an introduction to the author's contributions to the SOC area, resulting from his PhD research activity. It focuses on the problem of automatically composing a desired service, given a set of available ones and a target specification. As for description, services are represented as finite-state transition systems,  
so to provide an abstract account of their behavior, seen as the set of possible conversations with external clients. In addition, the presence of a finite shared memory is considered, that services can interact with and which provides a basic form of communication. Rather than describing technical details, we offer an informal overview of the whole work, and refer the reader to the original papers, referenced throughout this work, for all details.
\end{abstract}

\input{01-intro}
\bibliographystyle{eptcs} 
\bibliography{patrizi}

\end{document}

%% file: macros.tex
\newtheorem{example}{Example}[section]
\newtheorem{definition}{Definition}[section]

  {\trivlist\item[\hskip\labelsep\bf Proof:]}
  {\unskip\nobreak\hfill~$\Box$\endtrivlist}    

\begingroup
\catcode`\~=11
\gdef\urltilde{\lower 0.6ex\hbox{~}}
\endgroup

\newcommand{\C}{\mathcal{C}}

\renewcommand{\S}{\mathcal{S}}

\newcommand{\tlv}[0]{{\sc tlv}}
\newcommand{\colombo}[0]{{\sc Colombo}}

\newcommand{\smv}[0]{{\sc smv}}

\newcommand{\nop}[1]{{}}

\newcommand{\defterm}[1]{\mbox{\underline{\it\smash{#1}\vphantom{\lower.1ex\hbox
{x}}}}}

\newcommand{\lora}{\longrightarrow}

\newcommand{\goto}[1]{\stackrel{#1}{\lora}}

\newcommand{\tup}[1]{\langle #1\rangle}            

\newcommand{\Nat}{{\rm I\kern-.23em N}}

\newcommand{\myi}{\emph{(i)}\xspace}
\newcommand{\myii}{\emph{(ii)}\xspace}
\newcommand{\myiii}{\emph{(iii)}\xspace}



%% file: 01-intro.tex
\section{Introduction}
\nop{Services are software artifacts, possibly distributed and built on top of different technologies, that export a description of themselves, are accessible to external clients and communicate through a commonly known, standard, interface which enables inter-operability. More in general, Service Oriented Computing (SOC, cf., e.g., \cite{ACKM04}) is a computing paradigm whose basic elements are services, that can be used as building blocks to devise other services. A classical example of such paradigm is provided by Web Services, i.e., applications published over the Internet and self-described, usually built by different companies and relying on different technologies, which share a same communication protocol.~\footnote{SOAP built on top of HTTP.} For instance, online travel agencies such as Expedia~\footnote{\url{www.expedia.com}} integrate different web services offered by hotels, airlines, restaurants, etc., and provide final users with a complete service, combining all functionalities of its basic components. To do so, no constraint is required over the \emph{internal} structure of each web service, whereas they are all required to be published, compliant with the same communication protocol and to export a description of their interface, so to facilitate clients' access and communication. 

Abstracting from this example, services can be thought of as generic programs, publicly available and \emph{wrapped} so to mutually interact and communicate over a common platform (often referred to as \emph{middleware}). As such, the SOC paradigm makes easier code re-use and extension, as, in a sense, each service is interpreted as a usual \emph{method} in programming languages and, thus, a set of services, usually referred to as a \emph{community}, as a sort of \emph{programming library}. This similarity can be taken as the basis of \emph{service composition}: as exactly as in a programming language methods are combined to produce more complex methods, so services can be combined to build more complex services.
\medskip 
}
\noindent
This work provides an overview of author's contributions to the SOC area, resulting from his PhD research activity~\cite{patrizi-phd-09}, that can be summarized in the proposal of a novel technique for automated composition  of \emph{conversational} services, which is optimal (wrt time-complexity) and overcomes many of the obstacles encountered by similar existing proposals. The paper focuses on \emph{automated service composition}, that is, the problem of automatically combining a set of available services so as to meet a desired specification. Such a topic has a lot in common with other research areas, the most closely related being, probably, System Verification and Synthesis (e.g., \cite{PnRo89}), but also others provide theoretical frameworks that service composition can be cast over, such as Planning in Artificial Intelligence (e.g., \cite{pistore-etal-05}), Reasoning About Actions (e.g., \cite{mcilraith-son-02}) and even Data Integration (e.g., \cite{thakkar-etal-04}),  thus making available some research achievements that SOC research can benefit from. Due to this interdisciplinarity, several approaches have been proposed to model services, e.g., as atomic actions~\cite{mcilraith-son-02}, finite state machines~\cite{Hull@www03,berardi-etal-dl-03,BCDLM03c} or views over data~\cite{thakkar-etal-03}, and to solve the corresponding composition problem. Here, we follow the same approach as in~\cite{berardi-etal-dl-03,BCDLM03c} and adopt a \emph{behavioral model} of services. Starting from such work, we \myi consider some additional features, such as \emph{operation nondeterminism} and  \emph{presence of a shared memory} which allows for basic inter-service communication and, more importantly, \myii introduce a novel solution technique based on the formal notion of \emph{simulation relation}~\cite{Miln71}, which improves previous techniques in that it allows to compute the \emph{whole set of solutions}, at no additional (worst-case) computational time cost.
  
\subsection{Service Composition: An Overview}
Let us take a closer look at service composition, starting from the classical architecture for  Web Services.  Typically, the parties involved in a web service-based  session, besides the client, which can be a service itself, include two additional classes of entities, namely: a \emph{service broker} and some \emph{service providers}. The former is a central, well-known, registry which stores service descriptions and can be accessed by clients when searching for services that meet some desired requirements; the latter are organizations, such as companies, which make services actually available. Providers register their services to brokers and, when invoked, serve clients' requests. A typical session is as follows: \myi a client contacts one (or more) broker(s) and requests a service that meets a desired specification;  \myii if such a service is found then the broker refers the client to the service provider actually deploying the service; \myiii the client, which located the service, based on the broker's information, contacts the desired service and interacts with it.
\nop{
\begin{figure}
	\centering
	\includegraphics[width=.8\textwidth]{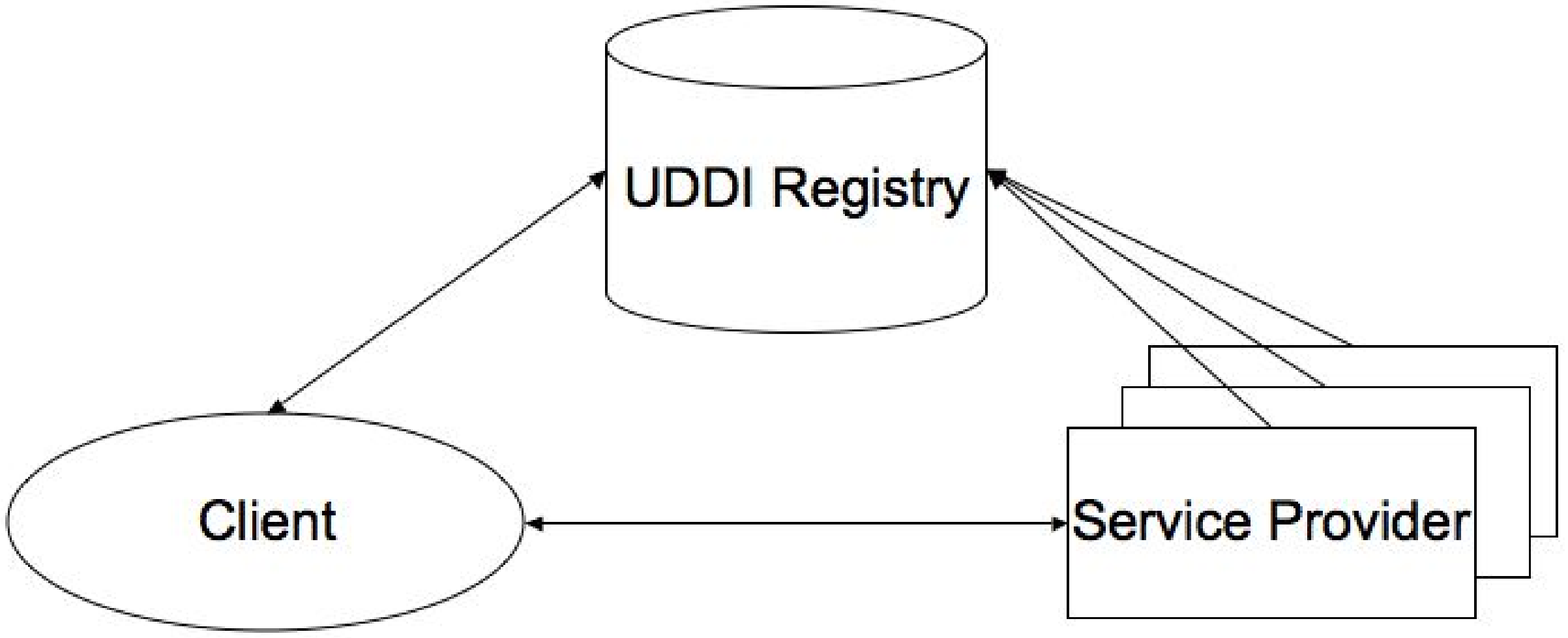}
\caption{Classical Web Service Architecture.\label{fig:wsoa}}
\end{figure}
}
Two classical questions about SOC arise:
\begin{enumerate}
	\item \emph{How are services described?}
	\item \emph{What if the desired service is not found?}  
\end{enumerate}
The first question concerns \emph{modeling}, i.e., the definition of a suitable service abstraction, able to capture aspects that can be relevant to client; the second one raises the problem of finding a constructive alternative to the trivial answer: ``client's request cannot be fulfilled, unless someone develops and deploys a new service that meets the desired requirements''. As one may expect, there exist many \emph{correct} answers to such questions. We address both problems. On the one hand, we propose a conceptual model (not an actual language), that substantially enriches existing ones (e.g.,~\cite{berardi-etal-08}), able to capture service \emph{behavior}, that is, which provides an abstraction of service evolution over time, representing their possible conversations with clients; on the other hand, on top of this model, we propose sound and complete techniques for \emph{building} a solution that fulfills a client's needs, when possible, by combining other available services. In particular, such techniques are shown to be the \emph{best one can do}, in that they return the most general solutions, while being optimal with respect to worst-case time complexity.
\section{Describing Service Behavior}
In the literature, several approaches to service modeling have been proposed. Rather than actual languages, such as WSDL, widely used to describe Web services, we focus on their \emph{conceptual model}. We can say that WSDL has an underlying \emph{atomic} conceptual model, specified in terms of input-output requirements. For instance, a service providing stock quotes of some market can be successfully described this way, with a single operation that returns the list of quotes. Such a model is useful in many situations, as its popularity over the Web witnesses, however, when more complex specifications need to be exported, it shows severe limitations. For instance, think of the same web service for stock quotes and assume that it provides quotations only to authenticated clients. In an input-output approach, one would describe two operations, say, \texttt{auth} and \texttt{quote}, as well as the respective data format necessary for interaction. Unfortunately, the input-output approach does not allow for \emph{conversation specification}, i.e., for putting constraints on the order that operations should be executed in. A very natural constraint would be, e.g., requiring clients to authenticate before requesting quotes. Observe also that cases may exist where two services export a same set of operations but allow for different execution sequences. Since this last constraint is not captured by input-output approaches, such services would appear to clients as the same one. In a word, atomic conceptual models export service \emph{interface} but not their \emph{behavior}.

The need for a \emph{behavioral} description of services is widely recognized (e.g., \cite{benatallah-etal-03,ACKM04,HBCS03,berardi-etal-dl-03}), yet, the community suffers from a lack of standard languages for this purpose. In this  work, we follow the same approach as the so-called Roman Model (\cite{Hull05}), originally introduced in~\cite{berardi-etal-dl-03,BCDLM03c}, which proposed the abstraction of conversational services as deterministic transition systems, where each state brings  information about both service's (relevant) past history and the potential (atomic) conversations that can be carried out with a client. Inspired by that, we propose a rich model, oriented to describe all \emph{conversations} supported by services, that includes relevant  features, such as nondeterminism and shared memory, and, thus, increases the set of actual scenarios that are captured. 

\begin{figure}
\centering
	\subfigure[][]{
		\label{subfig:wsdl-model}
		\includegraphics[width=.2\textwidth]{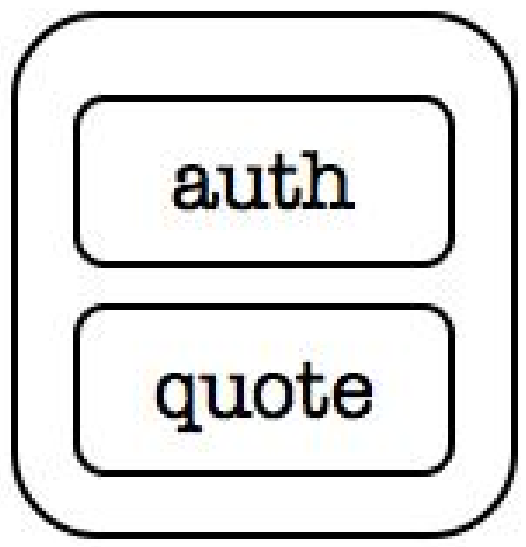}
	}
	\subfigure[][]{
		\label{subfig:beh-model}
		\includegraphics[width=.2\textwidth]{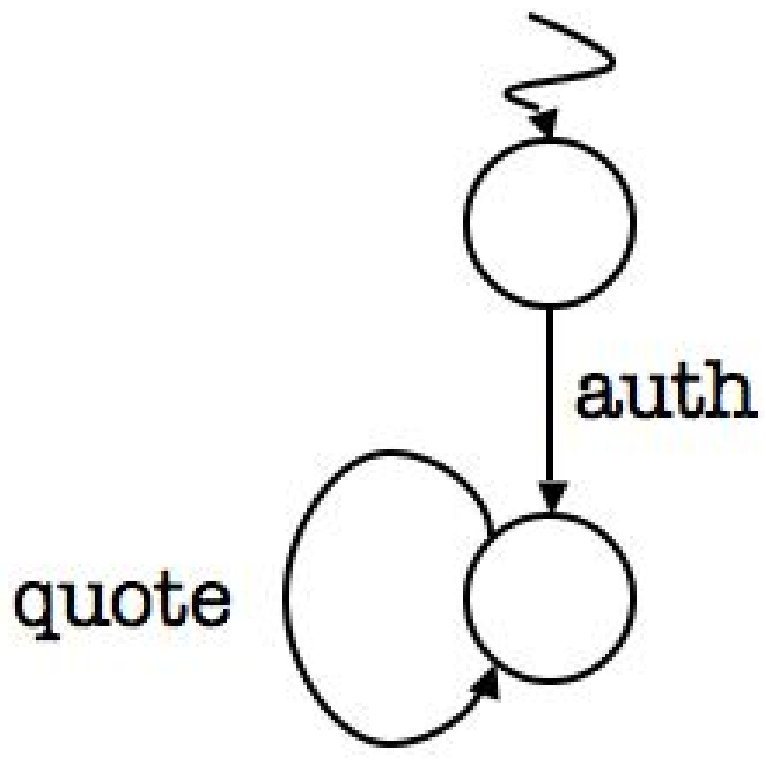}
	}
	\caption[Two approaches to service description]{Two approaches to service description: \ref{subfig:wsdl-model} Input-output model; \ref{subfig:beh-model} Behavioral model.}
\end{figure}

In our model, services export their behavioral specifications by means of an abstract language that represents transition systems, i.e., Kripke structures whose transitions are labeled by service operations, under the assumption that each legal run of the system corresponds to a conversation supported by the service. To clarify this, consider Figures~\ref{subfig:wsdl-model}~and~\ref{subfig:beh-model}. The former is a graphical representation of an input-output description of the stock quote service with authentication, as described above, which provides information about which operations can be requested; the latter is a behavioral representation of the same service, where more information is provided: indeed, it tells clients that they \myi must authenticate before requesting a \texttt{quote} operation and, then, \myii may request any number of quotes. Of course, more sophisticated examples do exist, where several operations, even nondeterministic, can be executed in a state, with nondeterminism modeling partial knowledge about service's internal logic. Also, there are settings relying on the same approach, where operations have parameters and are able to exchange data with other clients and even with an underlying database (cf. e.g., \cite{BCDHM05}).

A first advantage brought by such a model is its \emph{generality} with respect to service integration, in the sense that it is abstract enough to serve as conceptual model for several classes of scenarios. As an example, it can be used to model web service applications as well as multi-agent system ones. As a consequence, results obtained from this model are also relevant to areas different from SOC. Second, from the SOC viewpoint, it provides a behavioral, stateful, service representation, which allows for describing those inter-operation (temporal) constraints that current languages, e.g., WSDL, do not capture. We remark the importance of such a feature in a perspective of composition automatization: indeed, composition engines are intended to replace human operators, who compose services based on their informal description, often provided in natural language, which includes behavioral information.

Importantly, when dealing with a behavioral model, we can look at services as high-level descriptions of software artifacts. Indeed, they are characterized by states and state transitions triggered by inputs, which, specifically, represent requested operations. This interpretation suggests, hence, to see service (possibly finite) runs as computation fragments, that can be suitably combined to generate more complex services.

\section{Composing Services}
Many works exist which deal with automated composition of \emph{conversational} services, where  service  behavior is abstracted by various kinds of transition systems, e.g.,~\cite{terBeek-etal-07,berardi-05,Hull@www03,TrPi04}. The closest one to our work is ~\cite{berardi-05}, that we take as a starting point, where the problem of automatically composing a set of services, described as possibly nondeterministic transition systems, is addressed. Although here we propose a significant extension of~\cite{berardi-05}, in that we devise a novel solution technique which relies on effective technologies and yields great advantages, the basic problem has not changed. It can be informally stated as follows:
\begin{quote}
Consider a set of available services, a.k.a. \emph{community}, and an additional \emph{target service}, all exporting their conversational behavior. Is it possible to coordinate the available services so to support, at execution time, all conversations supported by the target service?
\end{quote}
In other words, the problem amounts to realize a (virtual) target service, by resorting only to (actual) available services. Obviously, how services are combined in the practice depends on the exported behavioral model. To see how this can be done under our model, consider the following example.
\begin{example}\label{ex:comp-ex}
Figure~\ref{fig:comp-ex} shows a service composition problem instance in the Roman Model, which includes two available services, represented in Subfigures~\ref{subfig:comp-ex-as1}~and~\ref{subfig:comp-ex-as2}, and a target one, in Subfigure~\ref{subfig:comp-ex-tgt}.
\begin{figure}
\centering
	\subfigure[][]{
		\label{subfig:comp-ex-as1}
		\includegraphics[width=.2\textwidth]{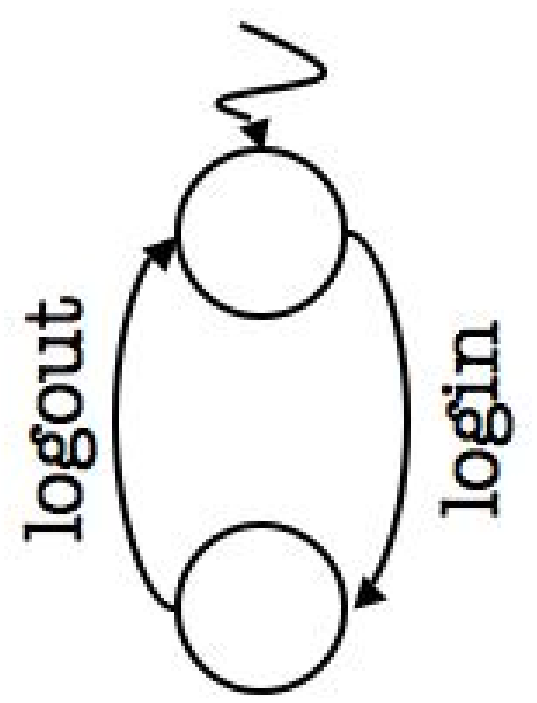}
	}
	\subfigure[][]{
		\label{subfig:comp-ex-as2}
		\includegraphics[height=.25\textwidth]{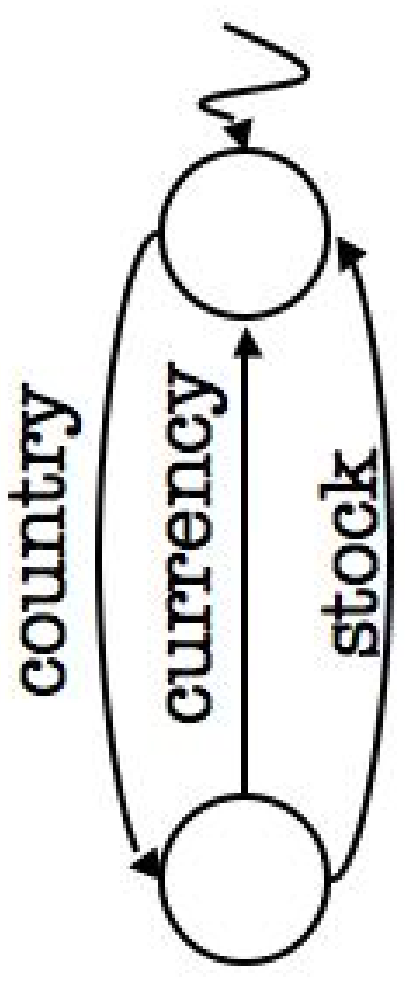}
	}
	\subfigure[][]{
		\label{subfig:comp-ex-tgt}
		\includegraphics[width=.4\textwidth]{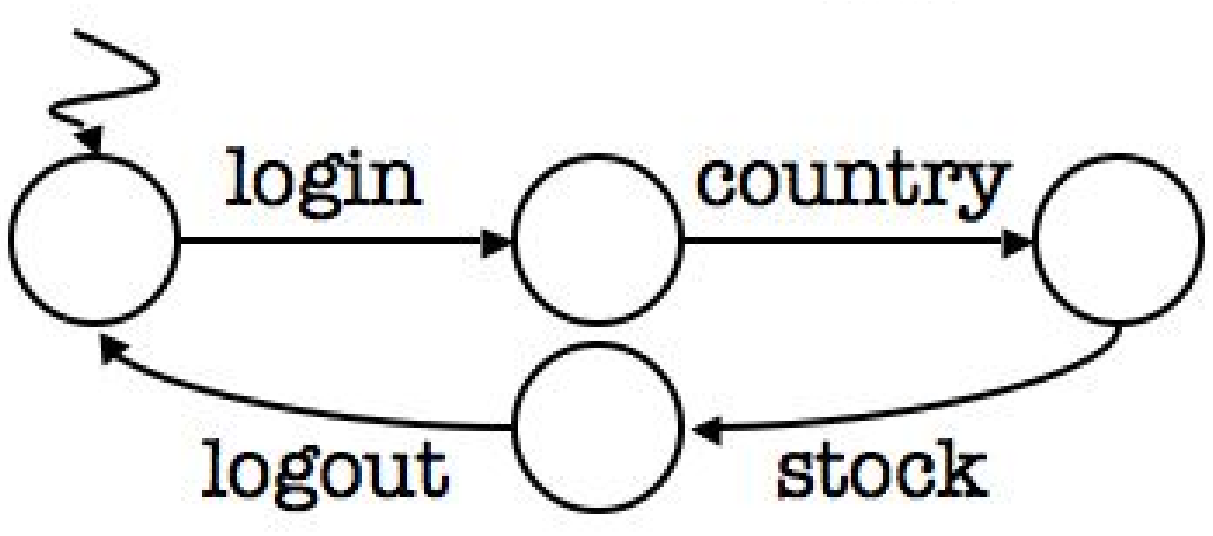}
	}
	\subfigure[][]{
		\label{subfig:comp-ex-comp}
		\includegraphics[width=.5\textwidth]{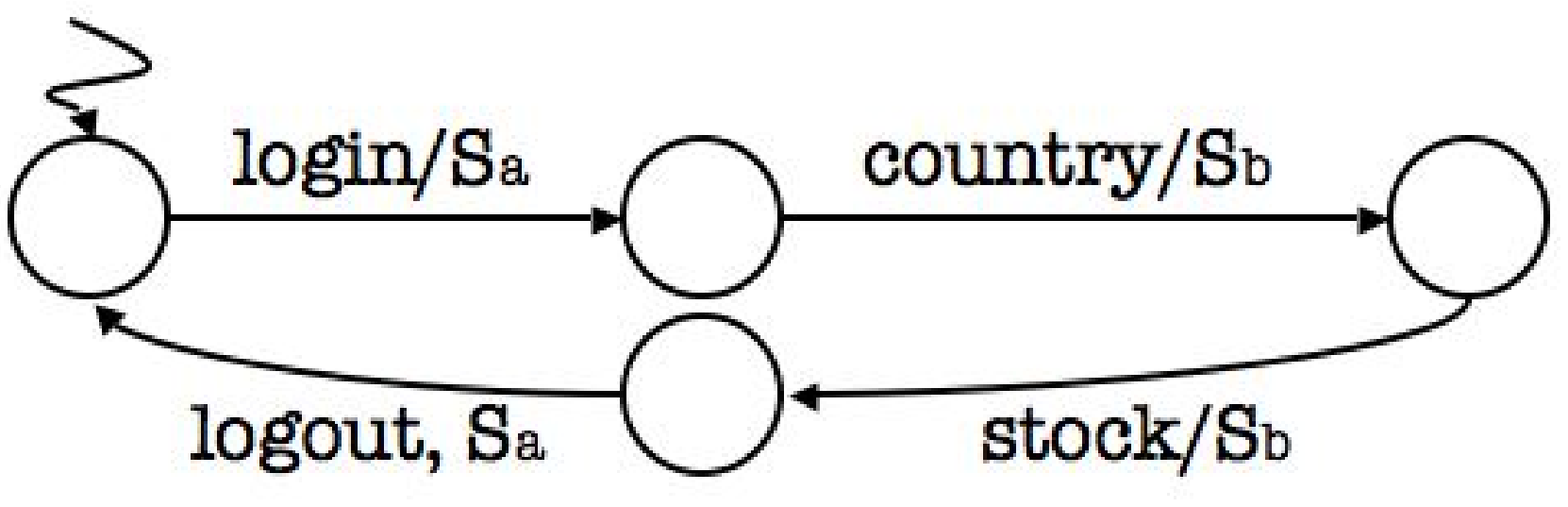}
	}
	\caption[A service composition example in the Roman Model]{A service composition example in the Roman Model\label{fig:comp-ex}: \ref{subfig:comp-ex-as1},\ref{subfig:comp-ex-as2} Available services; \ref{subfig:comp-ex-tgt} Target service; \ref{subfig:comp-ex-comp} A Composition.}
\end{figure}
The one in Subfigure~\ref{subfig:comp-ex-as1}, say $S_a$, provides login/logout capabilities, allowing a client to be authenticated and to close an authenticated session, whereas the one in Subfigure~\ref{subfig:comp-ex-as2}, say $S_b$, provides market stock quotes from all over the world. Clients willing to interact with $S_b$ are, first, required to input the market country of their interest and, then, are allowed to request either stock quotes or currency rates (versus, e.g., euro and dollar) for that market. As for the target service, say $S_c$, it provides stock quotes of a selected market only to authenticated clients. Specifically, clients of such service need first to login, then to select a market country, then are allowed to request quotes and, finally, to logout.

As we said, target services are \emph{virtual}, that is, only their specification exists, whereas their implementation is missing. However, it is easily seen that, by resorting to available services, this example's target service can be built. Indeed, it is enough \emph{delegating} login/logout operations to $S_a$ and country selection and stock requests to $S_b$. Observe that the target service not only provides a set of operations, but imposes a set of constraints over their executions, e.g., \texttt{stock} can be requested only after \texttt{country} has been executed. Since, on their side, also available service operations are subject to such kind of constraints, when a target service is to be realized, they must be met. For instance, had not $S_c$ required operation \texttt{country} be executed before \texttt{stock}, it would be not realizable, as $S_b$ is the only service that provides \texttt{stock} and it requires \texttt{country} to be executed first.
\end{example}
In Example~\ref{ex:comp-ex}, the composition can be realized by a machine which, on the one side, receives client operation requests and, on the other side, forwards them to an appropriate available service which executes them and, consequently, changes its state, where a new set of operations becomes available. Such a machine, similar to a Mealy machine but that can be, in general, infinite-state, is called an \emph{orchestrator}. One that solves the problem of Example~\ref{ex:comp-ex} is shown in Figure~\ref{subfig:comp-ex-comp}. Each state of the machine corresponds to a state of the target service and each transition is labeled by a pair of the form $operation/service$, with an intuitive semantics: the requested operation is assigned to the output service. For instance, operation \texttt{login} is delegated to service $S_a$.

The example above shows how the existence of temporal constraints among operation executions makes the problem non trivial: each time an operation is to be delegated to some available services, one needs to check whether all constraints are fulfilled, i.e., whether the service chosen for delegation is in a state where the operation is actually executable. This makes the orchestrator construction a hard task, akin to an advanced form of conditional planning~\cite{GhNTPlanning04}. Indeed, in the Roman Model, the service composition problem  is shown to be EXPTIME-complete~\cite{BCDLM03c,muscholl-walukiewicz-08}.

More complex scenarios can be considered. For instance, nondeterministic available services are also conceivable, where nondeterminism over operation execution represents partial knowledge about service's internal logic. Also, one could think of services communicating through a common blackboard or even exchanging data. All these scenarios require different notions of composition and, hence, different kind of orchestrators. \emph{The aim of our work is to provide a formal model for them and to propose a respective solution technique for the resulting composition problem}.

\bigskip\noindent
Before providing details about the techniques for composition problem solution, we mention a work~\cite{degiacomo-etal-07}, where the proposed technique applies to a more realistic scenario than those presented so far.  In particular, it shows how a workflow, to be carried out by a team of cooperating agents, is realized as coordination, or more precisely \emph{orchestration}, of several behaviors which provide high-level descriptions of agents' capabilities. Also, the approach is currently taken as a starting point for the development of a composition engine aimed at integrating embedded devices typically adopted in home automatization.~\footnote{This is part of recently started EU Project  SM4All (FP7-224332).}

\subsection{Solution Techniques}
Once the service composition problem and its solution have been defined, the problem of finding solution techniques that can be automated becomes central. 

\subsubsection{PDL-based solutions}
Previous work~\cite{berardi-05} addresses the problem of building a single orchestrator that is a solution to the problem. In such work, a technique was developed, able to deal with nondeterministic, finite-state, services, based on an encoding of the problem as a Propositional Dynamic Logic (PDL)~\cite{fischer-ladner-79} formula. Although such a technique is only able to build finite-state orchestrators, it is actually made effective by a crucial result showing that \emph{if an orchestrator exists then there exists one which is finite}~\cite{berardi-05}.

In a nutshell, PDLs constitute a family of logics that allow for specifying evolution of  propositional properties over time, in response to events. PDL models are Kripke structures, as often happens when dealing with dynamic systems. Importantly, for each PDL logic there exists an \emph{equivalent} Description Logic (DL), i.e., a logic used for \emph{static} knowledge representation, expressed in terms of \emph{classes} and \emph{relationships} among them~\cite{baader-etal-03}, such that each model of the former is a model of the latter and vice versa~\cite{schild-91,degiacomo-lenzerini-94}. So, since \myi PDLs capture the Roman Model and \myii effective DLs reasoning tools are available (e.g., {\sc FaCT}~\cite{horrocks-98}, {\sc Racer}~\cite{haarslev-moller-01}, Pellet~\footnote{\url{http://clarkparsia.com/pellet}}), one can exploit DLs to represent and actually solve  composition problems.

\begin{figure}
	\centering
	\includegraphics[width=.6\textwidth]{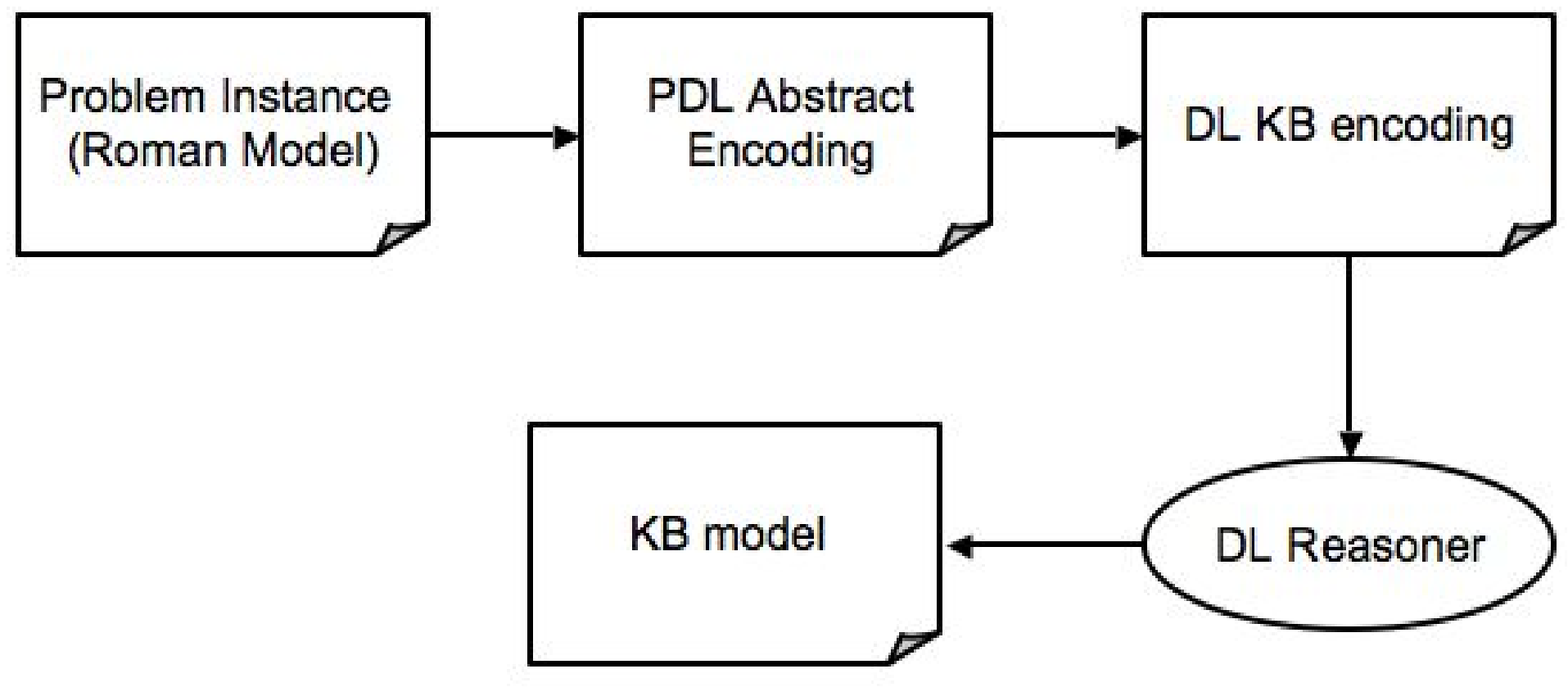}
\caption{Conceptual schema of PDL-based approach to service composition.\label{fig:pdl-solution}}
\end{figure}

A conceptual schema of the PDL-based approach adopted in~\cite{berardi-05} is depicted in Figure~\ref{fig:pdl-solution}.
Starting from a description of the problem, where all involved services are described by transition systems, first an abstract PDL formula is generated such that \myi each of its models (which are finite-state) corresponds to a (finite-state) orchestrator that is a solution to the original problem and, vice versa, \myii each composition problem's (finite-state)  solution has a corresponding model of the PDL-formula. Then, such a formula is translated into a DL knowledge base, represented in an actual format suitable for a DL reasoner which, finally, generates a model of the knowledge base, if consistent. By construction, such a model is also a model of the original PDL formula which, in turn, corresponds to a composition problem's solution. Based on this approach, an actual tool, $\mathcal{ESC}$ (E-Service Composer), has been devised (\cite{berardi-05}) which is able, with some limitations, to actually compute an orchestrator that realizes a target service. Unfortunately, this approach has three major drawbacks:
\begin{itemize}
	\item only finite-state orchestrators are returned;
	\item the obtained solution is not \emph{flexible}, that is, if a solution has been built which relies on an available service and such a service becomes unavailable at runtime, then the solution is no longer valid and the best one can do, using this approach, is to re-compute a new solution; 
	\item on the practical side, due to implemented DL reasoner limitations, $\mathcal{ESC}$ is actually able to \emph{synthesize} a model only for some particular inputs, though it is complete with respect to \emph{checking} for the existence of a model.
\end{itemize}
These limitations constitute the essential motivations to our work.

\subsubsection{Simulation-based solutions}
\begin{figure}
\centering
	\subfigure[][]{
		\label{subfig:simul-ex-ts2}
		\includegraphics[height=.3\textwidth]{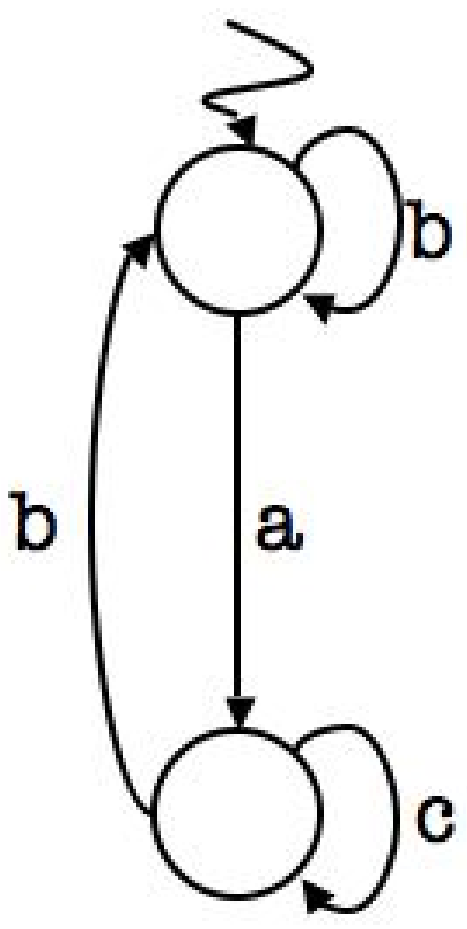}
	}
	\subfigure[][]{
		\label{subfig:simul-ex-ts1}
		\includegraphics[width=.2\textwidth]{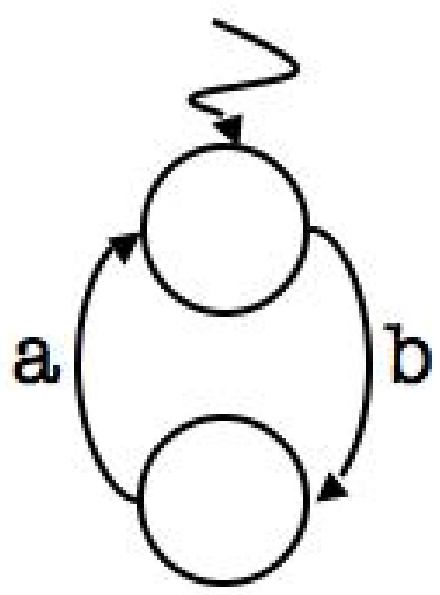}
	}
	\caption[Two transition systems]{Two transition systems.\label{fig:simul-ex}}
\end{figure}

We propose a novel solution technique based on the formal notion of  simulation relation between transition systems~\cite{Miln71}. Informally, given two transition systems $S_1$ and $S_2$, we say that $S_1$ \emph{simulates} $S_2$ if it shows, at least, the same behaviors as $S_2$. For example, considering Figure~\ref{fig:simul-ex}, assume that $S_1$ is the transition system shown in Subfigure~\ref{subfig:simul-ex-ts1} and $S_2$ is the one in Subfigure~\ref{subfig:simul-ex-ts2}. Seen as services,  $S_1$ simulates $S_2$, as each conversation supported by $S_2$, i.e., runs of the form $(\texttt{ab})^*$, is also supported by $S_1$. 
Of course, the vice versa does not hold: for instance, $(\texttt{b})^*$ is supported by $S_1$ but not by $S_2$. Although, here, we resort to regular languages as a means for describing service conversations, in our approach we do not adopt pure Finite State Machines (FSMs) as service models. Indeed, generic transition systems are better suited for our purposes, as we are interested in which \emph{choices} a service actually provides in each state. 

The following definition~\cite{Miln71}, where $S^f$ is used to identify the set of a system's final states, provides the central notion that our work is built upon. Actually, the one which follows can be used for our purposes only when dealing with deterministic transition systems, whereas for nondeterministic ones, a different notion, namely \emph{ND-simulation} is required. However, modulo formal details, the approach and the essential ideas remain the same.

\begin{definition}
Given two transition systems $\S_t$ and $\S_\C$, a simulation relation of $\S_t$ by $\S_\C$ is a relation $R\subseteq S_t\times S_\C$, such that:
\begin{itemize}
\item[] $\tup{s_t,s_\C}\in R$ implies:
       \begin{enumerate}
       		\item if $s_t\in S^f_t$ then $s_{\C}\in S^f_{\C}$;
       		\item for each transition $s_t \goto{o} s'_t$ in $\S_t$ there exists a transition $s_{\C} \goto{o} s_{\C}'$ in $\S_{\C}$ and $R(s'_t,s'_{\C})$.
        \end{enumerate}
\end{itemize}
\end{definition}
As it can be seen, this is a stronger relation than equivalence of FSMs, seen as language acceptors. Indeed, cases exist where two transition systems accept a same language but their states are not in simulation relation.

\nop{
For instance, consider the transition systems in Figure~\ref{fig:transition-systems}. As FSMs, they would be equivalent, since both
represent the language $\{ab, ac\}$, while, as Transition Systems (TSs),  they are indeed different:
\begin{itemize}
\item $TS_1$ is deterministic and models the case in which after operation $a$ one
  can perform both $b$ \emph{and} $c$;
\item $TS_2$ is nondeterministic and models the case in which, after operation $a$,
 one is allowed to perform either $b$ \emph{or} $c$, depending on the actual transition that takes place after executing $a$.
\end{itemize}

\begin{figure}
\centering
\subfigure[$TS_1$\label{subfig:ts1}]{
	\footnotesize
	\begin{pspicture}(0,0)(3,-4)
	\psgrid 
	\pnode(1.5,0){start}
	\cnodeput(1.5, -0.7){s0}{$s_0$} 
	\cnodeput(1.5, -2.2){s1}{$s_1$} 
	\cnodeput[doubleline=true](0.5, -3.6){s2}{$s_2$} 
	\cnodeput[doubleline=true](2.5, -3.6){s3}{$s_3$} 
	\ncline{->}{start}{s0}
	\ncline{->}{s0}{s1}\naput{$a$}
	\ncline{->}{s1}{s2}\nbput{$b$}	
	\ncline{->}{s1}{s3}\naput{$c$}
	\end{pspicture}
}
\subfigure[$TS_2$\label{subfig:ts2}]{
	\footnotesize
	\begin{pspicture}(0,0)(3,-4)
	\psgrid 
	\pnode(1.5,0){start}
	\cnodeput(1.5, -0.7){s0}{$s_0$} 
	\cnodeput(0.5, -2.2){s1}{$s_1$} 
	\cnodeput(2.5, -2.2){s2}{$s_2$} 
	\cnodeput[doubleline=true](0.5, -3.6){s3}{$s_3$} 
	\cnodeput[doubleline=true](2.5, -3.6){s4}{$s_4$} 
	\ncline{->}{start}{s0}
	\ncline{->}{s0}{s1}\nbput{$a$}
	\ncline{->}{s0}{s2}\naput{$a$}
	\ncline{->}{s1}{s3}\nbput{$b$}	
	\ncline{->}{s2}{s4}\naput{$c$}
	\end{pspicture}
}
\caption{Two different transition systems.\label{fig:transition-systems}}
\end{figure}
}

Essentially, our technique amounts to reducing the composition problem to the search for a simulation relation between the target service and the available service asynchronous product, which is itself a transition system. Precisely, it is the transition system which describes all the possible interleaved executions of available services or, in other words,  represents all \emph{potentialities} of the community services, seen as a whole. Recalling the above informal definition of a simulation relation, this corresponds to answering the question: \emph{``can available services be combined in order to include the same behaviors as the target service?''}, which is precisely our problem.

A fundamental advantage brought by our technique is that simulation-based solutions are, in fact, \emph{universal solutions}, or \emph{composition generators}, i.e., finite structures that represent \emph{all possible, even infinite-state, orchestrators that realize a target service}. Importantly, this does not affect worst-case time complexity: indeed, it is known that searching for a simulation between two transition systems is polynomial in the size of the systems, hence, since an asynchronous product result has exponential size with respect to its factor's size, it comes out that our technique requires exponential time with respect to the size of the (original) input systems. This, along with the observation that the service composition problem is proven EXPTIME-complete~\cite{muscholl-walukiewicz-08}, yields that also our simulation-based technique is optimal with respect to worst-case time complexity. In fact, with respect to the PDL-based approach, we obtain a complexity characterization refinement.

But our approach yields two additional benefits. 

First, by using composition generators, we obtain \emph{flexible} solutions, that is, able to change their behavior based on information available at runtime. The PDL-based approach does not provide this feature, since it returns a single, \emph{rigid}, solution that cannot be modified at runtime. For example if, during execution, an available service that the executing orchestrator relies on becomes unavailable, but can be replaced by another one, then with our solution we can change, without need for recomputation, the available service used to realize the target service. Differently put, we are able to \emph{switch} orchestrators during execution. 

\begin{figure}
	\centering
	\includegraphics[width=.7\textwidth]{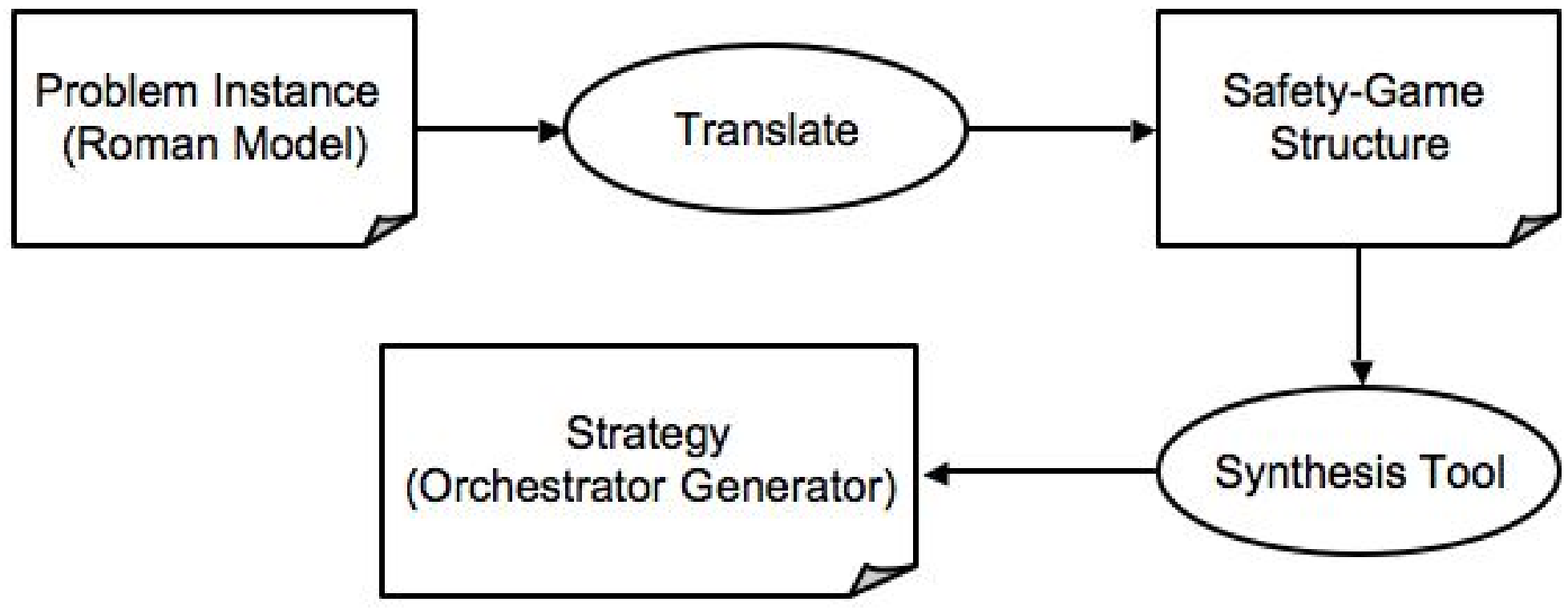}
\caption{Conceptual schema of the Game-based approach to service composition.\label{fig:game-solution}}
\end{figure}

Second, a set of effective tools, such as \tlv~\cite{pnueli-shahar-96}, Lily~\cite{jobstmann-etal-fmcad06} or Anzu~\cite{jobstmann-etal-07}, for computing simulation is becoming available. Precisely, such tools are aimed at synthesizing finite-state dynamic systems that meet desired temporal properties. In particular, they can be used to compute \emph{winning strategies} for \emph{safety games}, a.k.a. \emph{invariant} games, i.e., games where a player, in order to win, is required to maintain a given property all along game evolution. Indeed, our work proposes a reduction of the service composition problem into a safety game, such that computing a winning strategy for the obtained game corresponds to computing an orchestrator generator for the original problem, as defined in the simulation-based context. A conceptual schema of such an approach is depicted in Figure~\ref{fig:game-solution}. As synthesis engines are available, our efforts focused on defining the \emph{Translation} module, which implements a procedure for automatic reduction of a service composition instance into a game structure.
By doing so, we make available, for computing simulation relations, tools from the System Verification and Synthesis area, thus getting the major advantage of efficiency, as such tools resort to ordered binary decision diagrams (OBDD) for the internal representation of dynamic structures, thus limiting the typical state space explosion associated with synthesis procedures. 

\section{Further Issues}
In this Section, we briefly discuss additional issues which constitute interesting research directions  in service composition, that have been or are currently under investigation.

\subsection{Dealing with Data}
Data-Management capability is a desirable service property to capture: as a matter of fact, real services deal with data. Including such features in service representations would result in a more complete model, thus yielding the possibility of facing even more realistic problems. However, as several works witness, e.g., \cite{BCDHM05,deutsch-et-al-07,deutsch-etal-icdt09}, the presence of data poses a major obstacle to service composition and verification: data is infinite by its nature and makes services infinite-state. Unfortunately, the techniques described above, as well as others proposed in the literature, are effective only when dealing with finite-state systems, whereas infinite-state system verification and synthesis are known as a hard task for most non-trivial properties, and undecidable for general ones. Consequently, introducing data in service composition frameworks has a great impact on the solution techniques that can be adopted, thus making the problem a major challenge for the SOC community.

In our work, we propose a model that allows for dynamic and finite-state data structure representation~\footnote{Such a feature was first introduced in~\cite{IJCAI07}, in a different context.}. Precisely, in addition to available services, a community contains a so-called finite-state \emph{data box}, i.e., an additional  transition system modeling the evolution of a shared data structure whose state is affected by available services' execution, and that can be used to realize a basic form of inter-service communication. From a modeling perspective, a data box can be seen as a database with a finite-state behavior, used to capture some situations where services deal with data over finite domains. This way, we keep dealing with finite-state systems, thus making simulation-based techniques still applicable, but introduce a simple form of data-awareness. 

Such an extension, besides increasing the set of scenarios that can be captured and, hence, making the model appealing also for other areas~\footnote{Cf., e.g., \cite{sardina-etal-KR-08,degiacomo-etal-07} where multi-agent scenarios have been formulated as service composition ones},  provides the bases for more ambitious research, i.e., devising techniques for solving composition problem instances over generic data-aware services. The basic idea is \emph{abstracting} over actual data: first, a symbolic representation of actual database instances is built by using only a finite set of symbols; then, the infinite-state synchronous evolution of both the available services and the database is described as a finite-state system. This way, one can reduce the original problem to one defined over finite-state symbolic structures, thus making the above techniques still applicable. One attempt in this direction, in a synthesis context, is the \colombo\ framework described in~\cite{BCDHM05}, where services are able to exchange data from infinite domains and to interact with a relational database, by reading/writing scalar values. As for service verification, \cite{deutsch-et-al-07,deutsch-etal-icdt09} propose two similar abstraction techniques allowing for data-aware service verification.

We observe that~\cite{BCDHM05} tackles the composition problem by relying on a PDL-based approach. However, under the same model, one can recast the problem in terms of (data-aware) simulation, that is, by defining a relation between two data-aware services that interact with a common underlying data structure, whose data content may come from an infinite domain. This way, one would get the advantages brought by a simulation-based approach, though the actual resolution would be more complex, due to state space infiniteness, which calls for some abstraction procedure.
 
\subsection{Distributed Orchestrators}
In SOC, centralized scenarios, though conceivable, are rare. Even when all community services actually reside in a same repository, they communicate through common \emph{middleware} that enables communication and interoperability. Hence, the resulting abstraction is still a distributed community, where services are seen as programs, located at different places, that interoperate through some protocol. In such a distributed context, there can be cases where a central coordinating entity, such as an orchestrator, is not realizable nor convenient. In~\cite{AAAI07,degiacomo-etal-07}, two examples of such scenarios are reported. In particular, \cite{degiacomo-etal-07} shows one where all services execute on distributed peer devices, communicating via a wireless network. Clearly, in such situations one cannot rely on a single, central, coordinating unit: indeed, network disconnections are possible and can dramatically impact the overall system effectiveness. If, for instance, the coordinator permanently loses a connection, peers are no longer able to cooperate. Also less catastrophic events, such as temporary disconnections, may affect a system's efficiency and/or effectiveness. 

In our work, we built a peer-based framework for distributed orchestrator execution, that is, for execution in a scenario where each service has  an attached \emph{local} orchestrator, able to communicate with other ones, in order to cooperate for achieving a common goal.\\ 
As discussed above, orchestrators make their choices, i.e., select available services for delegation, based on community current  state, provided they have full observability on available service state. However, in this case, such assumption is no longer valid as, on the one hand, services reside at different places and, on the other hand and more importantly, there is no central entity able to \emph{observe} their state. In fact, local orchestrators have full observability only on the service they are attached to, and know nothing about other services. Therefore, even if a technique for distributed execution of a centralized orchestrator were given, since, in general, services are nondeterministic, a problem of \emph{state reconstruction} would arise. To solve this, we rely on the service communication ability, which is exploited to broadcast messages to other peers. Essentially, our technique amounts to building a usual orchestrator as if it were to be executed on a central coordinating device, and then, from this, generating one \emph{local orchestrator} for each available (distributed) service. Local orchestrators replicate a centralized one's behavior except for two features: \myi they are able to send/receive messages that are useful for community state reconstruction, i.e., they make their decisions based on received messages and current state of the service they are attached to, and \myii they issue commands only to the service they are attached to, while using received messages to \emph{keep track} of current community state. Differently put, each local orchestrator, through received messages, \emph{reconstructs} community current state and history until it is \emph{its turn}, and only then activates the respective service by issuing the command requested by the client.

This technique, along with the possibility of executing sevices in parallel, and not only under an interleaving policy, contributes to make the service-based approach quite general and, thus, appealing also for research areas other than SOC, e.g., Multi Agent Systems (MAS).

\section{Contributions}
The main goal of  our work is to address the drawbacks associated with a PDL-based solution approach to the service composition problem. In doing this, the achieved results introduce a set of novelties with respect to service composition state-of-the-art:
\begin{itemize}
	\item We propose a rich conceptual model for service behavioral description that makes services abstract enough to represent a variety of dynamic, possibly communicating, entities such as web services, physical devices or agents.
	\item We devise a novel approach to service composition that exploits the formal notion of simulation, thus bringing the following major advantages:
		\begin{itemize} 
			\item a huge literature about simulation, and simulation-related problems is available, as it is a standard notion, widely and very well studied in Computer Science and related areas, thus providing access to a set of useful results, such as efficient procedures for computing simulation relations (e.g.,~\cite{GePP03});
			\item the notion of simulation relation is associated with System Verification and Synthesis, thus making effective tools, based on Model Checking approach, available. In fact, we resort to one such tools to build our composition engine.
		\end{itemize}
	\item The proposed techniques overcome the obstacles met by the PDL-based approach: 
		\begin{itemize}
			\item  they return the whole (in general, infinite) set of solutions (including infinite-state ones), represented as a \emph{finite orchestrator generator}, without requiring additional computational efforts, with respect to previous approaches. This is undoubtedly the most relevant achievement, which shows the optimality of the approach.
			\item \emph{Just-in-time} solutions, able to adapt to run-time exceptional situations, can be built. With respect to the PDL-based approach, this corresponds to the possibility of \emph{switching} orchestrators during target service realization. Importantly, one can take advantage of information available at runtime, such as unexpected faults, which were considered \emph{catastrophic} in the previous approach;
			\item we get a refinement of the complexity bound obtained in~\cite{berardi-05}. Precisely, we identify  the exponential term as the number of available services, rather than their size.
		\end{itemize}
	\item We provide actual techniques that take advantage of flexible solutions to efficiently recompute an orchestrator generator when some exceptional situations arise, such as temporal/permanent unavailability of available services;
	\item We propose a formulation of the composition problem in a distributed scenario and show how it can be solved. The problem is not introduced to face a \emph{pure} service composition scenario, though it might be conceivable. Rather, it finds natural application in the Multi Agent System field, where several agents are described as possibly nondeterministic, communicating, finite-state transition systems (i.e., the same as services), operating in a common environment and communicating through a distributed shared memory;
	\item We show how the problem can be encoded into a safety-game structure, then written in an actual language, \smv, and, finally, processed by an implemented system, \tlv, to compute the whole set of solutions;
	\item Based on above game structure encoding, an actual system, Web Service Composition Engine (WSCE) was devised, which exploits \tlv\ for efficient problem solution.
\nop{
	\item We describe a feasible approach to tackle the problem of automatically composing data-aware services that interact with databases. In particular, we start from a \colombo-like framework and build a procedure to solve the problem when the underlying database has a bounded active domain that takes values from a totally ordered, dense, infinite domain. Although, on one side, such work has a lot of analogies with~\cite{BCDHM05}, on the other side, we propose an (ongoing) extension of our simulation-based approach to cope with operation parameters and databases,  bringing the advantages of \emph{flexibility} and \emph{robustness}, that typically come when this approach is adopted.
	}
\end{itemize}

%% file: patrizi.bbl
\begin{thebibliography}{10}
\providecommand{\bibitemstart}[1]{\bibitem{#1}}
\providecommand{\bibitemend}{}
\providecommand{\bibliographystart}{}
\providecommand{\bibliographyend}{}
\providecommand{\url}[1]{\texttt{#1}}
\providecommand{\urlprefix}{Available at }
\providecommand{\bibinfo}[2]{#2}
\bibliographystart

\bibitemstart{ACKM04}
\bibinfo{author}{G.~Alonso}, \bibinfo{author}{F.~Casati},
  \bibinfo{author}{H.~Kuno} \& \bibinfo{author}{V.~Machiraju}
  (\bibinfo{year}{2004}): \emph{\bibinfo{title}{{Web Services. Concepts,
  Architectures and Applications}}}.
\newblock \bibinfo{publisher}{Springer}.
\bibitemend

\bibitemstart{baader-etal-03}
\bibinfo{editor}{F.~Baader}, \bibinfo{editor}{D.~Calvanese},
  \bibinfo{editor}{D.L. McGuinness}, \bibinfo{editor}{D.~Nardi} \&
  \bibinfo{editor}{P.F. Patel-Schneider}, editors (\bibinfo{year}{2003}):
  \emph{\bibinfo{title}{The Description Logic Handbook: Theory, Implementation,
  and Applications}}. \bibinfo{publisher}{Cambridge University Press}.
\bibitemend

\bibitemstart{terBeek-etal-07}
\bibinfo{author}{M.H. ter Beek}, \bibinfo{author}{A.~Bucchiarone} \&
  \bibinfo{author}{S.~Gnesi} (\bibinfo{year}{2007}):
  \emph{\bibinfo{title}{{Formal Methods for Service Composition}}}.
\newblock {\sl \bibinfo{journal}{Annals of Mathematics, Computing and
  Teleinformatics}} \bibinfo{volume}{1}(\bibinfo{number}{5}), pp.
  \bibinfo{pages}{1--10}.
\bibitemend

\bibitemstart{benatallah-etal-03}
\bibinfo{author}{B.~Benatallah}, \bibinfo{author}{F.~Casati},
  \bibinfo{author}{F.~Toumani} \& \bibinfo{author}{R.~Hamadi}
  (\bibinfo{year}{2003}): \emph{\bibinfo{title}{{Conceptual Modeling of Web
  Service Conversations}}}.
\newblock In: {\sl \bibinfo{booktitle}{CAiSE}}. pp. \bibinfo{pages}{449--467}.
\bibitemend

\bibitemstart{berardi-05}
\bibinfo{author}{D.~Berardi} (\bibinfo{year}{2005}):
  \emph{\bibinfo{title}{Automatic Service Composition: Models, Techniques and
  Tools}}.
\newblock \bibinfo{type}{Ph.D. thesis}, \bibinfo{school}{{\sc Sapienza}
  Universit\`a degli Studi di Roma}.
\bibitemend

\bibitemstart{BCDHM05}
\bibinfo{author}{D.~Berardi}, \bibinfo{author}{D.~Calvanese},
  \bibinfo{author}{G.~De~Giacomo}, \bibinfo{author}{R.~Hull} \&
  \bibinfo{author}{M.~Mecella} (\bibinfo{year}{2005}):
  \emph{\bibinfo{title}{{Automatic Composition of Transition-based Semantic Web
  Services with Messaging}}}.
\newblock In: {\sl \bibinfo{booktitle}{Proc.\ of VLDB~2005}}.
\bibitemend

\bibitemstart{BCDLM03c}
\bibinfo{author}{D.~Berardi}, \bibinfo{author}{D.~Calvanese},
  \bibinfo{author}{G.~{De Giacomo}}, \bibinfo{author}{M.~Lenzerini} \&
  \bibinfo{author}{M.~Mecella} (\bibinfo{year}{2003}):
  \emph{\bibinfo{title}{{Automatic Composition of e-{S}ervices that Export
  their Behavior}}}.
\newblock In: {\sl \bibinfo{booktitle}{Proc.\ of ICSOC~2003}}. pp.
  \bibinfo{pages}{43--58}.
\bibitemend

\bibitemstart{berardi-etal-dl-03}
\bibinfo{author}{D.~Berardi}, \bibinfo{author}{D.~Calvanese},
  \bibinfo{author}{G.~De~Giacomo}, \bibinfo{author}{M.~Lenzerini} \&
  \bibinfo{author}{M.~Mecella} (\bibinfo{year}{2003}):
  \emph{\bibinfo{title}{{e-Service Composition by Description Logics Based
  Reasoning}}}.
\newblock In: {\sl \bibinfo{booktitle}{Description Logics}}.
\bibitemend

\bibitemstart{berardi-etal-08}
\bibinfo{author}{D.~Berardi}, \bibinfo{author}{F.~Cheikh},
  \bibinfo{author}{G.~{De Giacomo}} \& \bibinfo{author}{F.~Patrizi}
  (\bibinfo{year}{2008}): \emph{\bibinfo{title}{Automatic Service Composition
  via Simulation}}.
\newblock {\sl \bibinfo{journal}{International Journal of Foundations of
  Computer Science}} \bibinfo{volume}{19}(\bibinfo{number}{2}), pp.
  \bibinfo{pages}{429--451}.
\bibitemend

\bibitemstart{Hull@www03}
\bibinfo{author}{T.~Bultan}, \bibinfo{author}{X.~Fu}, \bibinfo{author}{R.~Hull}
  \& \bibinfo{author}{J.~Su} (\bibinfo{year}{2003}):
  \emph{\bibinfo{title}{{Conversation Specification: A New Approach to Design
  and Analysis of E-Service Composition}}}.
\newblock In: {\sl \bibinfo{booktitle}{Proc.\ of WWW~2003}}.
\bibitemend

\bibitemstart{degiacomo-lenzerini-94}
\bibinfo{author}{G.~De~Giacomo} \& \bibinfo{author}{M.~Lenzerini}
  (\bibinfo{year}{1994}): \emph{\bibinfo{title}{Boosting the Correspondence
  between Description Logics and Propositional Dynamic Logics}}.
\newblock In: {\sl \bibinfo{booktitle}{AAAI}}. pp. \bibinfo{pages}{205--212}.
\bibitemend

\bibitemstart{degiacomo-etal-07}
\bibinfo{author}{G.~{De Giacomo}}, \bibinfo{author}{M.~de~Leoni},
  \bibinfo{author}{M.~Mecella} \& \bibinfo{author}{F.~Patrizi}
  (\bibinfo{year}{2007}): \emph{\bibinfo{title}{{Automatic Workflows
  Composition of Mobile Services}}}.
\newblock In: {\sl \bibinfo{booktitle}{ICWS}}. pp. \bibinfo{pages}{823--830}.
\bibitemend

\bibitemstart{IJCAI07}
\bibinfo{author}{G.~De~Giacomo} \& \bibinfo{author}{S.~Sardi{\~n}a}
  (\bibinfo{year}{2007}): \emph{\bibinfo{title}{Automatic Synthesis of New
  Behaviors from a Library of Available Behaviors}}.
\newblock In: {\sl \bibinfo{booktitle}{Proc.\ of IJCAI~2007}}. pp.
  \bibinfo{pages}{1866--1871}.
\bibitemend

\bibitemstart{deutsch-etal-icdt09}
\bibinfo{author}{A.~Deutsch}, \bibinfo{author}{R.~Hull},
  \bibinfo{author}{F.~Patrizi} \& \bibinfo{author}{V.~Vianu}
  (\bibinfo{year}{2009}): \emph{\bibinfo{title}{{Automatic Verification of
  Data-Centric Business Processes}}}.
\newblock In: {\sl \bibinfo{booktitle}{Proc.\ of ICDT~2009}}.
\bibitemend

\bibitemstart{deutsch-et-al-07}
\bibinfo{author}{A.~Deutsch}, \bibinfo{author}{L.~Sui} \&
  \bibinfo{author}{V.~Vianu} (\bibinfo{year}{2007}):
  \emph{\bibinfo{title}{Specification and verification of data-driven Web
  applications}}.
\newblock {\sl \bibinfo{journal}{J. Comput. Syst. Sci.}}
  \bibinfo{volume}{73}(\bibinfo{number}{3}), pp. \bibinfo{pages}{442--474}.
\newblock \urlprefix\url{http://dx.doi.org/10.1016/j.jcss.2006.10.006}.
\bibitemend

\bibitemstart{fischer-ladner-79}
\bibinfo{author}{M.J. Fischer} \& \bibinfo{author}{R.E. Ladner}
  (\bibinfo{year}{1979}): \emph{\bibinfo{title}{Propositional Dynamic Logic of
  Regular Programs}}.
\newblock {\sl \bibinfo{journal}{J. Comput. Syst. Sci.}}
  \bibinfo{volume}{18}(\bibinfo{number}{2}), pp. \bibinfo{pages}{194--211}.
\bibitemend

\bibitemstart{GePP03}
\bibinfo{author}{R.~Gentilini}, \bibinfo{author}{C.~Piazza} \&
  \bibinfo{author}{A.~Policriti} (\bibinfo{year}{2003}):
  \emph{\bibinfo{title}{From Bisimulation to Simulation: Coarsest Partition
  Problems.}}
\newblock {\sl \bibinfo{journal}{J. Autom. Reasoning}}
  \bibinfo{volume}{31}(\bibinfo{number}{1}), pp. \bibinfo{pages}{73--103}.
\bibitemend

\bibitemstart{GhNTPlanning04}
\bibinfo{author}{M.~Ghallab}, \bibinfo{author}{D.~Nau} \&
  \bibinfo{author}{P.~Traverso} (\bibinfo{year}{2004}):
  \emph{\bibinfo{title}{Automated Planning: Theory and Practice}}.
\newblock \bibinfo{publisher}{Morgan Kauffman}.
\bibitemend

\bibitemstart{haarslev-moller-01}
\bibinfo{author}{V.~Haarslev} \& \bibinfo{author}{R.~M{\"o}ller}
  (\bibinfo{year}{2001}): \emph{\bibinfo{title}{{Description of the RACER
  System and its Applications}}}.
\newblock In: {\sl \bibinfo{booktitle}{Description Logics}}.
\bibitemend

\bibitemstart{horrocks-98}
\bibinfo{author}{I.~Horrocks} (\bibinfo{year}{1998}):
  \emph{\bibinfo{title}{{The FaCT System}}}.
\newblock In: {\sl \bibinfo{booktitle}{TABLEAUX}}. pp.
  \bibinfo{pages}{307--312}.
\bibitemend

\bibitemstart{Hull05}
\bibinfo{author}{R.~Hull} (\bibinfo{year}{2005}): \emph{\bibinfo{title}{Web
  Services Composition: A Story of Models, Automata, and Logics}}.
\newblock In: {\sl \bibinfo{booktitle}{2005 IEEE International Conference on
  Services (SCC 2005)}}.
\bibitemend

\bibitemstart{HBCS03}
\bibinfo{author}{R.~Hull}, \bibinfo{author}{M.~Benedikt},
  \bibinfo{author}{V.~Christophides} \& \bibinfo{author}{J.~Su}
  (\bibinfo{year}{2003}): \emph{\bibinfo{title}{{E-Services: a Look Behind the
  Curtain}}}.
\newblock In: {\sl \bibinfo{booktitle}{Proc.\ of PODS~2003}}. pp.
  \bibinfo{pages}{1--14}.
\bibitemend

\bibitemstart{jobstmann-etal-fmcad06}
\bibinfo{author}{B.~Jobstmann} \& \bibinfo{author}{R.~Bloem}
  (\bibinfo{year}{2006}): \emph{\bibinfo{title}{Optimizations for {LTL}
  Synthesis}}.
\newblock In: {\sl \bibinfo{booktitle}{Proc. of FMCAD '06}}.
  \bibinfo{publisher}{IEEE Computer Society}, \bibinfo{address}{Washington, DC,
  USA}, pp. \bibinfo{pages}{117--124}.
\bibitemend

\bibitemstart{jobstmann-etal-07}
\bibinfo{author}{B.~Jobstmann}, \bibinfo{author}{S.~Galler},
  \bibinfo{author}{M.~Weiglhofer} \& \bibinfo{author}{R.~Bloem}
  (\bibinfo{year}{2007}): \emph{\bibinfo{title}{Anzu: A Tool for Property
  Synthesis}}.
\newblock In: {\sl \bibinfo{booktitle}{Proc.\ of CAV~2007}}. pp.
  \bibinfo{pages}{258--262}.
\newblock \urlprefix\url{http://dx.doi.org/10.1007/978-3-540-73368-3_29}.
\bibitemend

\bibitemstart{mcilraith-son-02}
\bibinfo{author}{S.A. McIlraith} \& \bibinfo{author}{T.C. Son}
  (\bibinfo{year}{2002}): \emph{\bibinfo{title}{{Adapting Golog for Composition
  of Semantic Web Services}}}.
\newblock In: {\sl \bibinfo{booktitle}{KR}}. pp. \bibinfo{pages}{482--496}.
\bibitemend

\bibitemstart{Miln71}
\bibinfo{author}{R.~Milner} (\bibinfo{year}{1971}): \emph{\bibinfo{title}{An
  Algebraic Definition of Simulation Between Programs.}}
\newblock In: {\sl \bibinfo{booktitle}{Proc.\ of IJCAI~1971}}. pp.
  \bibinfo{pages}{481--489}.
\bibitemend

\bibitemstart{muscholl-walukiewicz-08}
\bibinfo{author}{A.~Muscholl} \& \bibinfo{author}{I.~Walukiewicz}
  (\bibinfo{year}{2008}): \emph{\bibinfo{title}{{A lower bound on web services
  composition}}}.
\newblock {\sl \bibinfo{journal}{Logical Methods in Computer Science}}
  \bibinfo{volume}{4}(\bibinfo{number}{2}).
\bibitemend

\bibitemstart{patrizi-phd-09}
\bibinfo{author}{F.~Patrizi} (\bibinfo{year}{2009}):
  \emph{\bibinfo{title}{{Simulation-Based Techniques for Automated Service
  Composition}}}.
\newblock \bibinfo{type}{Ph.D. thesis}, \bibinfo{school}{{\sc Sapienza}
  Universit\`a degli Studi di Roma}.
\bibitemend

\bibitemstart{pistore-etal-05}
\bibinfo{author}{M.~Pistore}, \bibinfo{author}{P.~Traverso} \&
  \bibinfo{author}{P.~Bertoli} (\bibinfo{year}{2005}):
  \emph{\bibinfo{title}{Automated Composition of Web Services by Planning in
  Asynchronous Domains}}.
\newblock In: {\sl \bibinfo{booktitle}{Proc.\ of ICAPS~2005}}. pp.
  \bibinfo{pages}{2--11}.
\bibitemend

\bibitemstart{PnRo89}
\bibinfo{author}{A.~Pnueli} \& \bibinfo{author}{R.~Rosner}
  (\bibinfo{year}{1989}): \emph{\bibinfo{title}{{On the Synthesis of a Reactive
  Module}}}.
\newblock In: {\sl \bibinfo{booktitle}{Proc.\ of POPL~1989}}. pp.
  \bibinfo{pages}{179--190}.
\bibitemend

\bibitemstart{pnueli-shahar-96}
\bibinfo{author}{A.~Pnueli} \& \bibinfo{author}{E.~Shahar}
  (\bibinfo{year}{1996}).
\newblock \emph{\bibinfo{title}{The {TLV} system and its applications}}.
\newblock \bibinfo{howpublished}{Technical report, Weizmann Institute}.
\bibitemend

\bibitemstart{sardina-etal-KR-08}
\bibinfo{author}{S.~Sardi{\~n}a}, \bibinfo{author}{G.~De~Giacomo} \&
  \bibinfo{author}{F.~Patrizi} (\bibinfo{year}{2008}):
  \emph{\bibinfo{title}{Behavior Composition in the Presence of Failure}}.
\newblock In: {\sl \bibinfo{booktitle}{Proc. of KR'08}}.
\bibitemend

\bibitemstart{AAAI07}
\bibinfo{author}{S.~Sardi{\~n}a}, \bibinfo{author}{F.~Patrizi} \&
  \bibinfo{author}{G.~{De Giacomo}} (\bibinfo{year}{2007}):
  \emph{\bibinfo{title}{Automatic Synthesis of a Global Behavior from Multiple
  Distributed Behaviors.}}
\newblock In: {\sl \bibinfo{booktitle}{Proc.\ of AAAI~2007}}. pp.
  \bibinfo{pages}{1063--1069}.
\bibitemend

\bibitemstart{schild-91}
\bibinfo{author}{K.~Schild} (\bibinfo{year}{1991}): \emph{\bibinfo{title}{A
  Correspondence Theory for Terminological Logics: Preliminary Report}}.
\newblock In: {\sl \bibinfo{booktitle}{IJCAI}}. pp. \bibinfo{pages}{466--471}.
\bibitemend

\bibitemstart{thakkar-etal-04}
\bibinfo{author}{S.~Thakkar}, \bibinfo{author}{J.L. Ambite} \&
  \bibinfo{author}{C.A. Knoblock} (\bibinfo{year}{2004}):
  \emph{\bibinfo{title}{A data integration approach to automatically composing
  and optimizing web services}}.
\newblock In: {\sl \bibinfo{booktitle}{In Proceedings of the ICAPS Workshop on
  Planning and Scheduling for Web and Grid Services}}.
\bibitemend

\bibitemstart{thakkar-etal-03}
\bibinfo{author}{S.~Thakkar}, \bibinfo{author}{C.A. Knoblock} \&
  \bibinfo{author}{J.L. Ambite} (\bibinfo{year}{2003}): \emph{\bibinfo{title}{A
  View Integration Approach to Dynamic Composition of Web Services}}.
\newblock In: {\sl \bibinfo{booktitle}{Proc. of ICAPS'03 Workshop on Planning
  for Web Services}}.
\bibitemend

\bibitemstart{TrPi04}
\bibinfo{author}{P.~Traverso} \& \bibinfo{author}{M.~Pistore}
  (\bibinfo{year}{2004}): \emph{\bibinfo{title}{Automated Composition of
  Semantic Web Services into Executable Processes}}.
\newblock In: {\sl \bibinfo{booktitle}{Proc.\ of ISWC-04}}, {\sl
  \bibinfo{series}{LNCS}} \bibinfo{volume}{3298}.
  \bibinfo{publisher}{Springer}, pp. \bibinfo{pages}{380--394}.
\bibitemend

\bibliographyend
\end{thebibliography}
